\documentclass[showpacs,aps,PRL,twocolumn,nofootinbib,11]{revtex4-1}

\usepackage{graphicx}
\usepackage{amssymb} 
\usepackage{amsmath} 
\usepackage{times}

\usepackage{txfonts}
\usepackage{lineno}

 \usepackage{ifpdf}
 \ifpdf
 \usepackage[pdftex]{hyperref}
 \else
 \usepackage[hypertex]{hyperref}
 \fi

 \hypersetup{
   pdftitle={},%
   pdfauthor={},%
   pdfsubject={},%
   pdfkeywords={},%
   pdfstartview={},%
   bookmarksopen=true, breaklinks=true, debug=true, %
   colorlinks=true, linkcolor=red, citecolor=blue, urlcolor=blue
 }

\usepackage{float} 
\usepackage{subfig} 

\usepackage{amsmath}
\usepackage{epsfig}
\usepackage{float}
\usepackage{verbatim} 
\usepackage{color} 

\newcommand{\ep}{\varepsilon}

\newcommand{\eqs}[1]{\begin{equation} \begin{split} #1\end{split} \end{equation} }

\newcommand{\ie}{{\it i.e.}}
\newcommand{\eg}{{\it e.g.}}

\newcommand{\Q}{{\cal Q}}
\def\so{{\bigl.^3\hspace{-1mm}S^{[8]}_1}}
\def\mopb{{\langle\mathcal{O}(\bigl.^3\hspace{-1mm}S_1^{[8]})\rangle}}

\newcommand{\cf}[1]{{Fig.~\ref{#1}}}

\begin{document}

\title{Production of $\boldsymbol{J/\psi+\eta_c}$ vs. $\boldsymbol{J/\psi+J/\psi}$  at the LHC: Importance of the Real $\boldsymbol{\alpha_s^5}$ Corrections}

\author{Jean-Philippe Lansberg$^a$, Hua-Sheng Shao$^{b,c}$}

\affiliation
{ 
$^a$ IPNO, Universit\'e Paris-Sud, CNRS/IN2P3, F-91406, Orsay, France \\
$^b$ Department of Physics and State Key Laboratory of Nuclear Physics and Technology,
Peking University, Beijing 100871, China\\
$^c$ PH Department,TH Unit,CERN,CH-1211 Geneva 23,Switzerland
}

\begin{abstract}
We proceed for the first time to the evaluation of the Born cross section for 
$J/\psi+\eta_c$ production, namely via $g+g \to J/\psi+\eta_c+g$, and show that 
it has a harder $P_T$ spectrum  than the $J/\psi$-pair yield at Born level. If 
one stuck to a comparison at Born level, one would conclude that $J/\psi+\eta_c$ 
production would surpass that of $J/\psi+J/\psi$ at large $P_T$. This is nonetheless 
not the case since $J/\psi$-pair production, as for single $J/\psi$, receives 
leading-$P_T$ contributions at higher orders in $\alpha_s$. We also present the first evaluation
of these leading-$P_T$ next-to-leading order contributions. These are indeed significant for increasing $P_T$ and 
are of essential relevance for comparison with forthcoming data. 
We also compute kinematic correlations relevant for double-parton-scattering studies.
Finally, we evaluate the polarisation of a $J/\psi$ accompanied either by a $\eta_c$ or a $J/\psi$
 and another light parton. These results may be of great help to understand the polarisation of 
quarkonia produced at high energies. 
\end{abstract}

\maketitle

{\it Introduction.---}Since a long time, quarkonium physics has reached a precision era. Yet, 
a number of puzzles still challenge our understanding of their production mechanism 
(see~\cite{Lansberg:2006dh,Brambilla:2010cs,ConesadelValle:2011fw} for reviews),
hence of QCD at the interplay between its short- and long-distance domains.

With the advent of the LHC, data at higher energies, at higher transverse 
momenta, with higher precision and with more exclusivity towards direct 
production are now flowing in. Unfortunately, all this may not be sufficient 
to pin down the complexity of the quarkonium-production mechanism. In this context, 
much hope is put into the study of associated-quarkonium 
production, in particular that of a pair of quarkonia.

We know that $\alpha^4_s$ and $\alpha^5_s$ corrections to the colour-singlet 
mechanism (CSM)~\cite{CSM_hadron} are significant and cannot be overlooked if one tries to 
explain the $P_T$ dependence of the $J/\psi$ and $\Upsilon$ cross sections observed in 
high-energy hadron collisions~\cite{Campbell:2007ws,Artoisenet:2007xi,Gong:2008sn,Gong:2008hk,Artoisenet:2008fc,Lansberg:2008gk}. 
As far as the $P_T$-integrated yield is concerned, 
the CSM contributions agree relatively well with the existing data at colliders
energies~\cite{Brodsky:2009cf,Lansberg:2010cn}. It is thus natural to wonder 
whether this also applies to quarkonium-pair production.

Polarisation predictions are also dramatically affected by QCD corrections, both in the inclusive 
case and in the production of quarkonia with a prompt 
photon~\cite{Gong:2008sn,Gong:2008hk,Artoisenet:2008fc,Lansberg:2010vq,Li:2008ym,Lansberg:2009db}. 
First 3-D analyses of the $J/\psi$ and $\psi(2S)$ polarisation have been recently carried out at the LHC by the 
ALICE~\cite{Abelev:2011md}, CMS~\cite{Chatrchyan:2013cla} and LHCb~\cite{Aaij:2013nlm} collaborations. They reveal an 
unexpected unpolarised yield in disagreement with 
basically all the available models. It is therefore also expedient to look at the polarisation pattern of quarkonia
produced in pairs.

Recently, LHCb has studied two associated production channels of $J/\psi$:  
$J/\psi$ + charm~\cite{Aaij:2012dz} and a pair of $J/\psi$~\cite{Aaij:2011yc}. 
The latter process has been measured for the first time by the CERN-NA3 collaboration in the eighties
but at large $x_F$~\cite{Badier:1982ae,Badier:1985ri}. The rates were higher than 
expected and seemed to be only explained  by the coalescence of 
double intrinsic charm pair in the proton projectile~\cite{Vogt:1995tf}. The cross
 section measured by LHCb covers a totally different region, 
which a priori should be accounted for by the conventional pQCD approaches, \eg~by the 
CSM. Indeed, the $P_T$-integrated rate obtained by LHCb are in very good agreement 
with the recent theoretical expectations from the CSM~\cite{Qiao:2009kg,Berezhnoy:2011xy} 
based on the pioneer works of Kartvelishvili \& Esakiya~\cite{Kartvelishvili:1984ur} and 
Humpert~\cite{Humpert:1983yj}. It thus seems that, as far as the $P_T$-integrated yields 
are concerned, the CSM predictions are as good for single inclusive as for double 
inclusive $J/\psi$ production.

Because of $C$-parity conservation, it is believed that the $J/\psi+\eta_c$ yield, 
as well as $J/\psi+\chi_c$, will be suppressed. It has even been suggested~\cite{Berezhnoy:2012xq}
that the expected relative suppression between $J/\psi+\chi_c$ and $J/\psi+J/\psi$ 
could provide a handle to extract the double-parton-scattering (DPS)
contributions as opposed to the single-parton scatterings (SPS) evaluated here. The 
former could indeed be a significant source of quarkonium pairs at the LHC~\cite{Kom:2011bd}.

Inclusive $\eta_c$ production is certainly difficult to study experimentally. 
Attempts are being made by the LHCb collaboration to
look at them in the $p\bar p$ decay channel along with the other charmonia~\cite{Barsuk:2012ic}. 
Searches in such  decay channels as well as in some of the dominant 3/4 body decays, 
\eg~ $\eta_c \to K\bar K\pi^0$ or $\eta_c \to K^+K^-\pi^+\pi^-$, pose significant challenges 
in terms of background and 
triggering. A possible option to partially circumvent these difficulties may be to 
search for $\eta_c$ in the existing samples of $J/\psi$ to look for $J/\psi+\eta_c$.

Our motivation was therefore to evaluate the yield for $J/\psi+\eta_c$ at leading 
order and to compare it to that of $J/\psi$ pair production. For a fair comparison 
of the $P_T$ dependence, we have also evaluated --for the first time-- the leading-$P_T$ 
NLO contributions to the latter production. Our evaluation consists in the 
computation the real-emission NLO corrections regulated by a cut-off (NLO$^\star$) instead of by the loop 
corrections in a full NLO computation. To enrich the comparison, we have also computed
some kinematic correlations whose analysis may help to study the DPS contributions and, finally, 
the polarisation of a $J/\psi$ associated with a $\eta_c$ and of a $J/\psi$ in a pair of $J/\psi$
taking into account the leading-$P_T$ NLO QCD corrections.

\begin{figure}[h!]
\centering
\subfloat[]{\includegraphics[width=.25\columnwidth,draft=false]{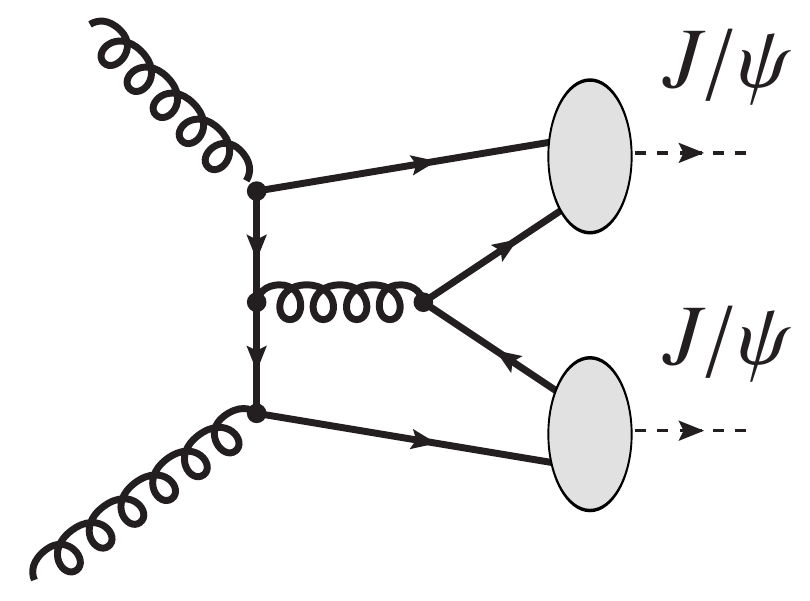}\label{diagram-a1}}
\subfloat[]{\includegraphics[width=.25\columnwidth,draft=false]{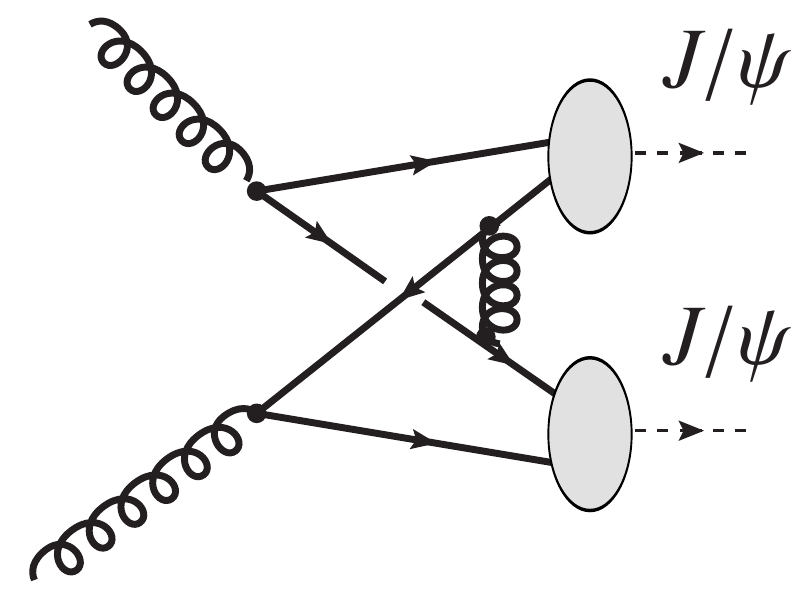}\label{diagram-a2}}
\subfloat[]{\includegraphics[width=.25\columnwidth,draft=false]{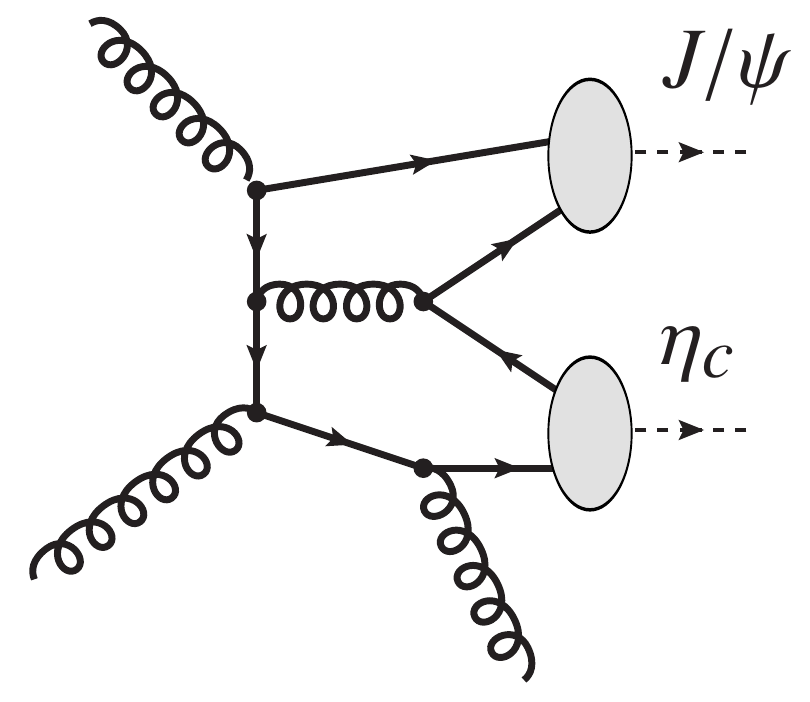}\label{diagram-b}}
\subfloat[]{\includegraphics[width=.25\columnwidth,draft=false]{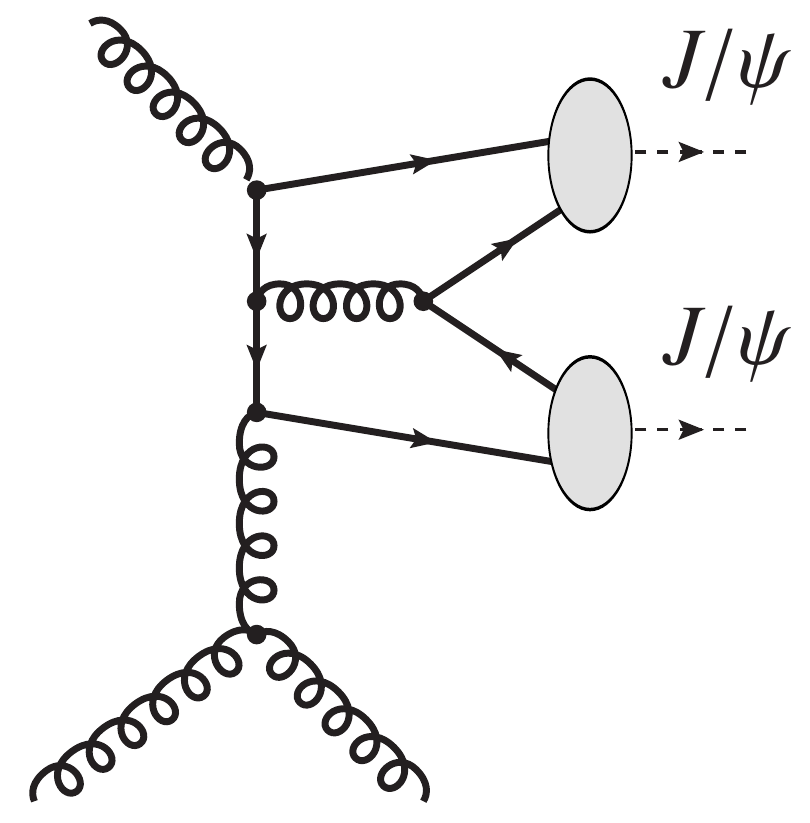}\label{diagram-c}}
\caption{Some diagrams contributing to the hadroproduction of a pair of quarkonium $\Q+ \Q'$ in 
the CSM at orders $\alpha_S^4$ (a \& b) for $J/\psi + J/\psi$, and $\alpha_S^5$ for $J/\psi + \eta_c$ (c) and 
for $J/\psi +  J/\psi$ (d). The quarks and antiquarks attached to the ellipses are taken 
as on-shell and their relative velocity $v$ is set to zero.}
\label{diagrams}
\end{figure}

{\it Cross-section  evaluation .---}
In the CSM~\cite{CSM_hadron}, the amplitude for the production of a pair of $S$-wave 
quarkonia ${\Q_1}$ and  ${\Q_2}$, of given momenta $P_{1,2}$ and of polarisation $\lambda_{1,2}$ 
accompanied by other partons, noted $k$, is written as the product of the amplitude to create 
the corresponding double heavy-quark pair, in each of which the relative momentum of the heavy 
quarks ($p_{1,2}$) is zero, two spin projectors $N(\lambda_{1,2}| s_{1,3},s_{2,4})$ and $R_{1,2}(0)$, 
the radial wave functions at the origin in the configuration space for both quarkonia. 
Precisely, one has
\eqs{\label{CSMderiv3}
{\cal M}(ab \to &{\Q_1}^{\lambda_1}(P_1)+{\Q_2}^{\lambda_2}(P_2)+k)=\\\!\sum_{s_1,s_2,c_1,c_2}&\sum_{s_3,s_4,c_3,c_4}\!\!\frac{N(\lambda_1| s_1,s_2)N(\lambda_2| s_3,s_4)}{ \sqrt{m_{\Q_1}m_{\Q_2}}} \frac{\delta^{c_1c_2}\delta^{c_3c_4}}{N_c}\frac{R_1(0)R_2(0)}{4 \pi}\\\times& {\cal M}(ab \to Q^{s_1}_{c_1} \bar Q^{s_2}_{c_2}(\mathbf{p_1}=\mathbf{0}) + Q^{s_3}_{c_3} \bar Q^{s_4}_{c_4}(\mathbf{p_2}=\mathbf{0}) + k),\nonumber}
where one defines from the heavy-quark momenta, $q_{1,2,3,4}$, 
$P_{1,2}=q_{1,3}+q_{2,4}$, $p_{1,2}=(q_{1,3}-q_{2,4})/2$, and where $s_{1,3}$,$s_{2,4}$ are the 
heavy-quark spin components and $\delta^{c_ic_j}/\sqrt{N_c}$  is the CS projector. $N(\lambda| s_i,s_j)$ 
is the spin projector, which has a simple expression in the non-relativistic limit: 
$\frac{1}{2 \sqrt{2} m_{Q} } \bar{v} (\frac{\mathbf{P}}{2},s_j) \Gamma_{S} u (\frac{\mathbf{P}}{2},s_i) \,\, $ 
where $\Gamma_S$ is $\gamma_5$ when $S=0$ (\eg~ $\eta_c$), and $\ep^{\lambda}_{\mu}\gamma^{\mu}$ when $S=1$ 
(\eg~ $J/\psi$). In this analysis, we will use the generic tree-level matrix-element and event generator for heavy 
quarkonia, HELAC-Onia, described in~\cite{Shao:2012iz}, to perform all the calculations.  We have considered the
processes of $J/\psi$-pair and $J/\psi+\eta_c$ production at LO via $gg\to J/\psi+J/\psi$ at $\alpha_s^4$ 
(see \eg~\cf{diagram-a1}-\ref{diagram-a2}) and $gg\to J/\psi+\eta_c+g$ at $\alpha_s^5$ (see \eg~\cf{diagram-b}). 
In the latter case, the emission of a final state gluon is imposed by $C$-parity conservation and the gluon is 
necessarily radiated by the heavy-quark line. Note that the quark-induced processes do not contribute at these orders.

In addition, we have studied the impact of the real-emission QCD corrections to direct $J/\psi$-pair 
production (see \eg~\cf{diagram-c}). 
The contribution of these added to that at LO is what we call the NLO$^\star$ yield.
In order to avoid infrared divergences (IR) appearing when real-emissions are considered, we 
followed~\cite{Artoisenet:2008fc} and have imposed that the invariant mass of any light-parton pair, $s_{ij}$, 
be larger than the IR cut-off $s_{ij}^{\rm{min}}$. 

Our IR treatment is expected to give a reliable estimation of the NLO result at 
least at large $P_T$ --and probably at mid $P_T$-- for the following reasons. 
One notes first that, by inspection of all propagators, 
one can easily see that our IR cut-off, $s_{ij}>s_{ij}^{\rm{min}}$, is sufficient 
to regulate all the collinear and soft divergences in the real-emission corrections 
to $J/\psi$-pair production.
The key argument is then that for the new $P_T$-enhanced topologies appearing at NLO, from \eg~the 
$t$-channel-exchange diagram shown in Fig.\ref{diagram-c}, $s_{ij}$ will necessarily be large for any 
light-parton pair at large $P_T$. For the remaining topologies, one may encounter large logarithms of
$s_{ij}/s_{ij}^{\rm{min}}$, but this are factors of the amplitudes of $P_T$-suppressed topologies and
the dependence on $s_{ij}^{\rm{min}}$ should therefore vanishes as soon as $P_T$ increases. 
Finally, we note that the 
virtual corrections, which necessarily have the same $P_T$-scaling as the Born contributions, 
are also $P_T$ suppressed compared to the real-emission contributions and they can also be 
neglected since we have anyhow regulated the IR divergences.

The reliability of the NLO$^\star$ approximation has been explicitly verified in the 
inclusive $J/\psi$ and $\Upsilon$ production~\cite{Artoisenet:2008fc,Lansberg:2008gk}, 
as well as for $J/\psi+\gamma$ and $\Upsilon+\gamma$~\cite{Lansberg:2009db}\footnote{It is 
also worth noting that the NLO$^\star$ of $J/\psi+Z$ production 
reproduces quite well the exact NLO result~\cite{Gong:2012ah}, though there is 
another large energy scale $m_Z$ besides $P_T$.}. As we show in the next section, 
the scaling of the NLO$^\star$ yield for $J/\psi+J/\psi$ is clearly enhanced by $P_T^2$ 
compared to the LO yield and the IR cut-off sensitivity vanish extremely quickly. These are
clear indications that the method works for this process at this order.

As regards the parameters entering the computations, we have taken $|R_{J/\psi,\eta_c}(0)|^2=0.81$~GeV$^3$ and
$M_{J/\psi,\eta_c}=2m_c$. Our uncertainty bands are obtained from the {\it combined} 
variations of $m_c=1.5\pm 0.1$ GeV, with the factorisation $\mu_F$ and the renormalisation $\mu_R$ 
scales chosen among the couples $(0.5 \mu_0, 2 \mu_0)$, where $\mu_0=m_T=\sqrt{(4m_c)^2+p_T^2}$.

\begin{figure*}[ht!]
\begin{center}
\subfloat[central $y$]{\includegraphics[width=0.33\textwidth,draft=false]{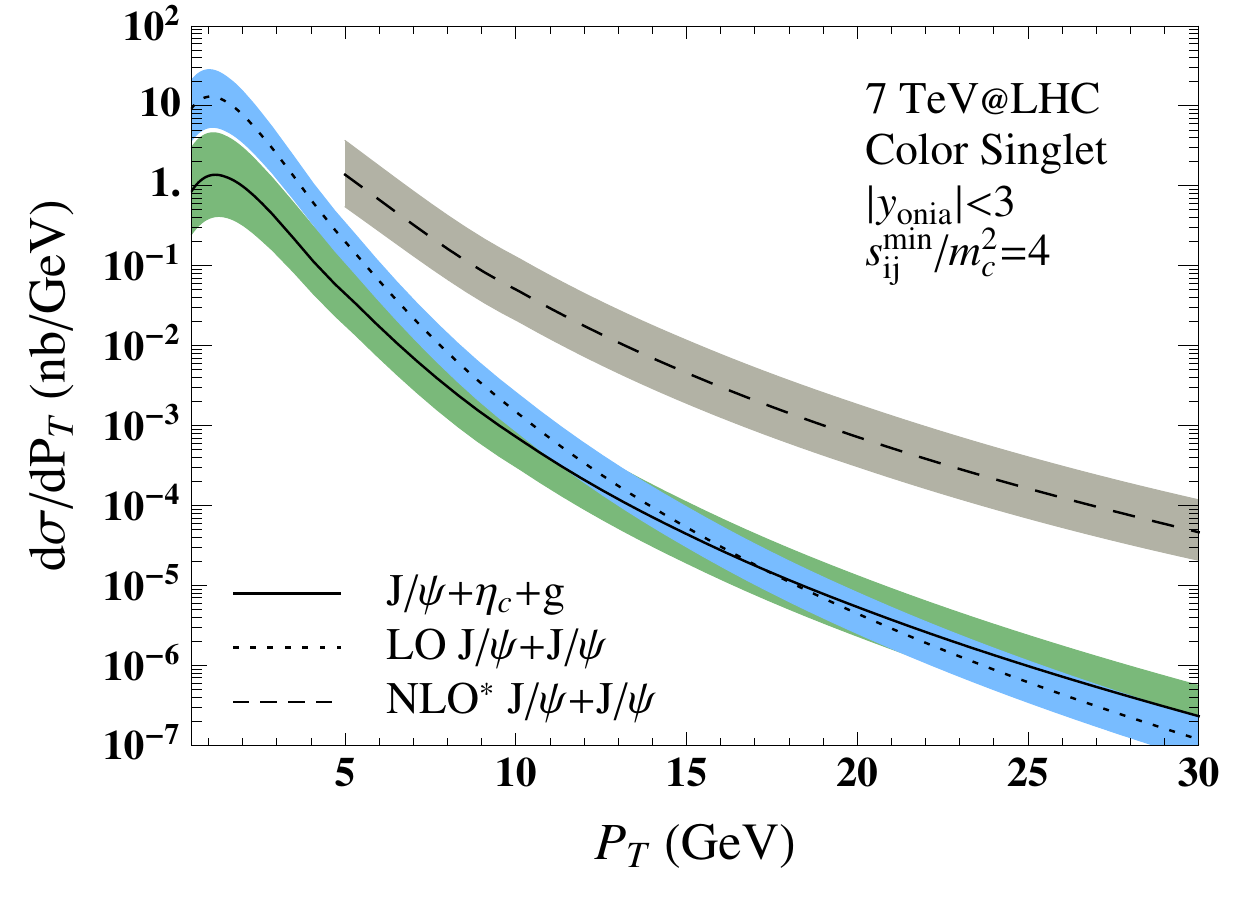}\label{fig:dsigdPTa}}
\subfloat[forward $y$ ]{\includegraphics[width=0.33\textwidth,draft=false]{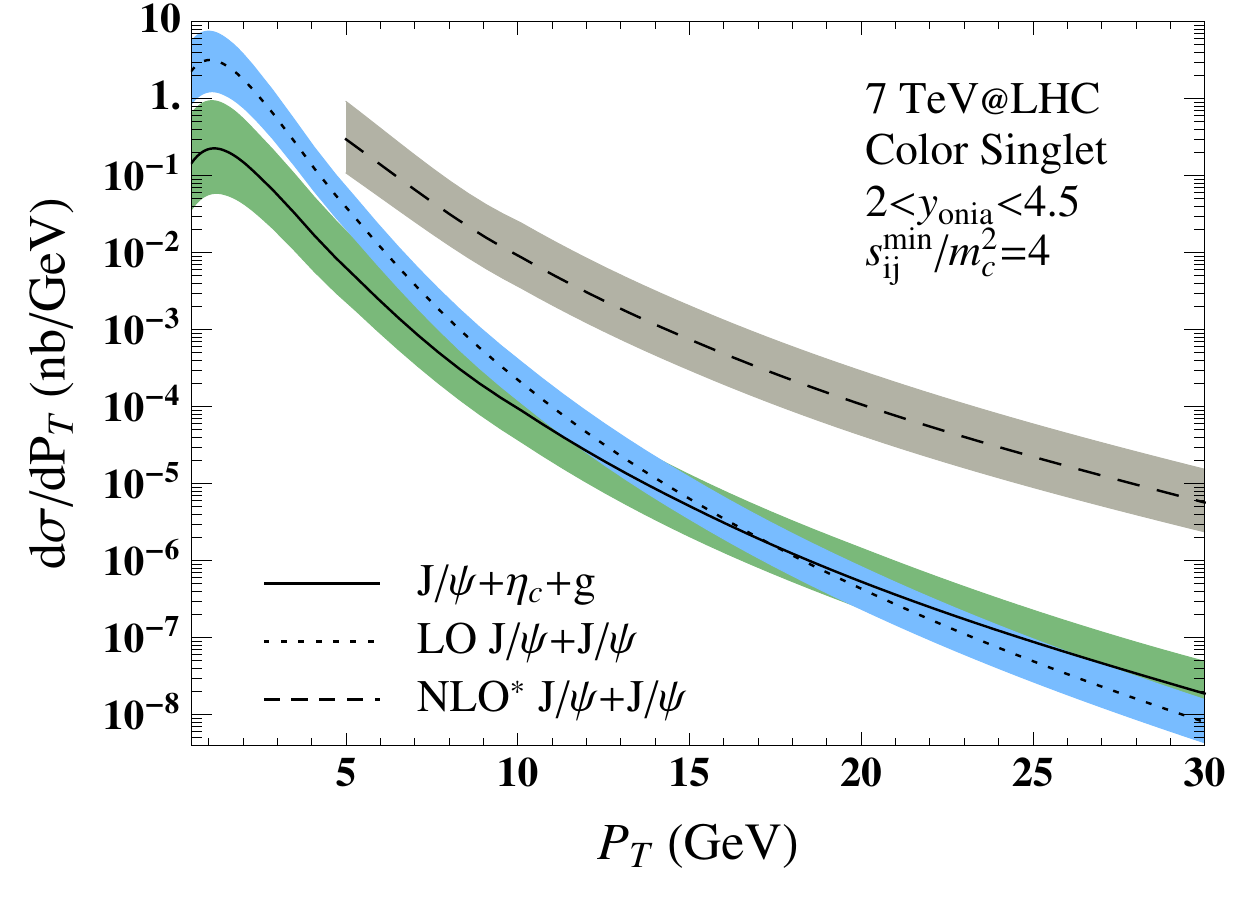}\label{fig:dsigdPTb}}
\subfloat[Comparison with COM and $s_{ij}^{\rm min}$ sensitivity]{\includegraphics[width=0.33\textwidth,draft=false]{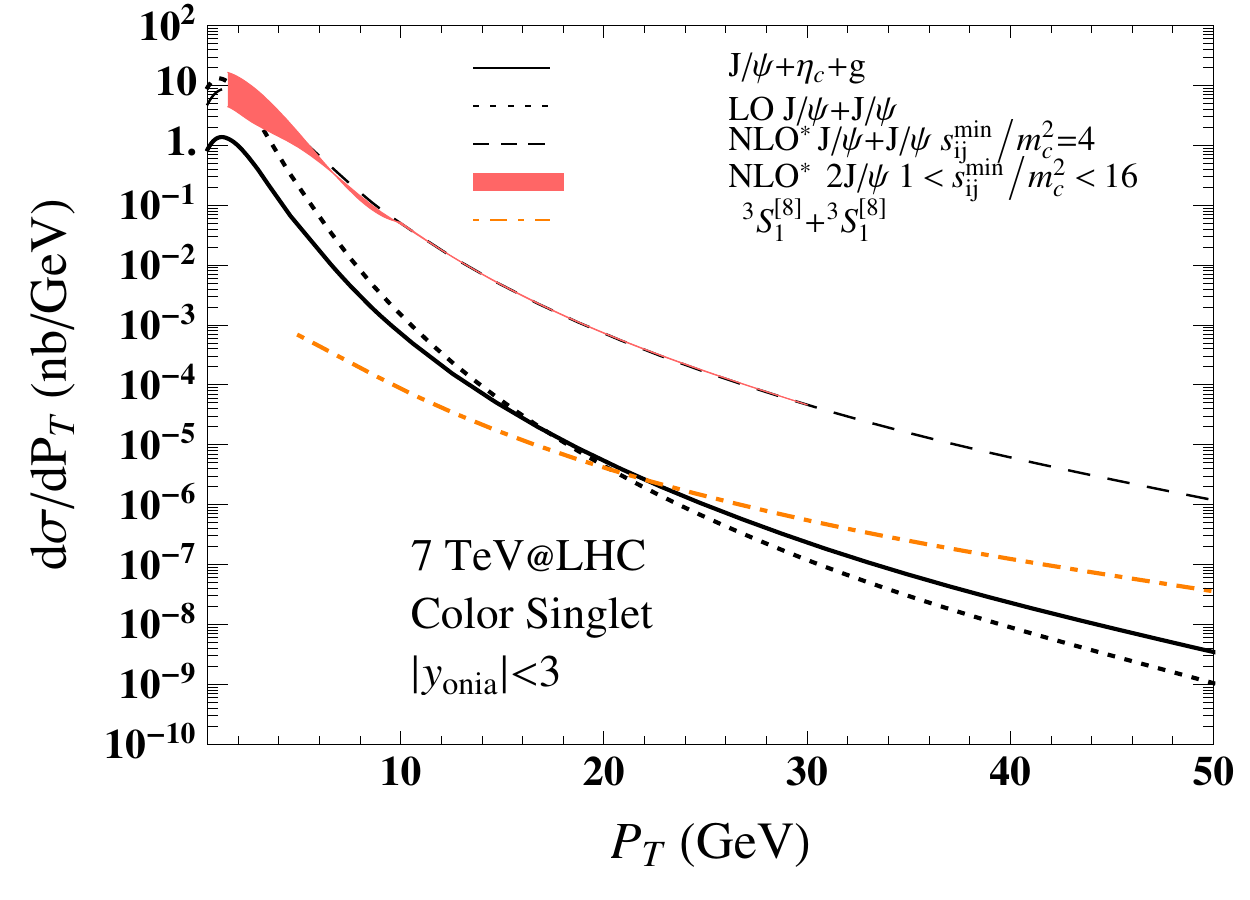}\label{sijdep}}
\caption{$d\sigma/dP_T$ at 7 TeV for $J/\psi+J/\psi$ at LO and NLO$^\star$ (+ COM),  and for $J/\psi+\eta_c+g$}
\label{fig:dsigdPT}
\end{center}
\end{figure*}

\begin{figure*}[ht!]
\vspace*{-1cm}
\begin{center}
\subfloat[Azimuthal correlations]{
\includegraphics[trim = 0mm 0mm 0mm 0mm,clip,height=4.445cm,draft=false]{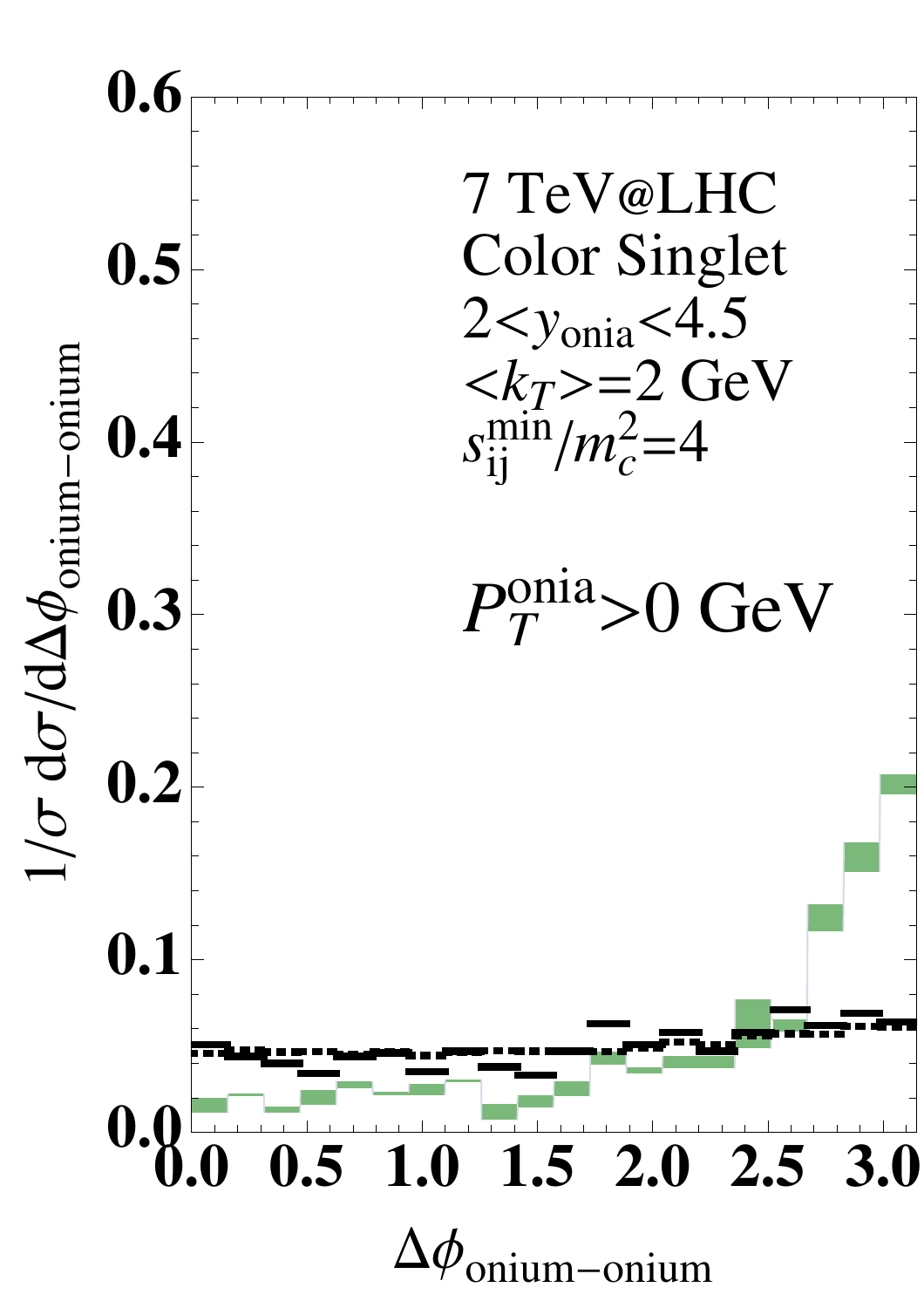}
\includegraphics[trim = 21.8mm 0mm 0mm 0mm,clip,height=4.42cm,draft=false]{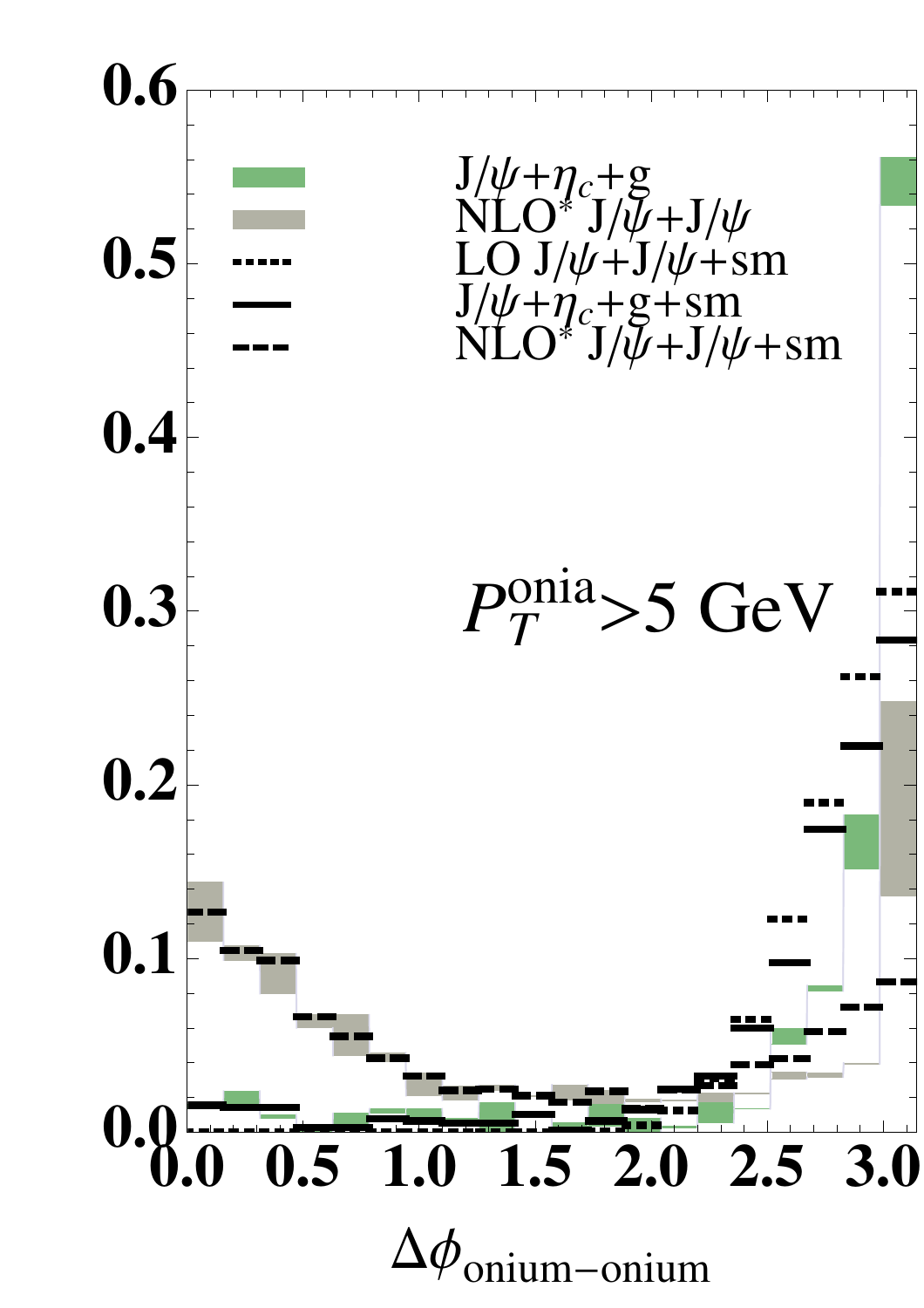}\label{fig:dphi}}\hspace*{1mm}
\subfloat[Normalised invariant-mass distributions]{\includegraphics[height=4.2cm,clip,draft=false]{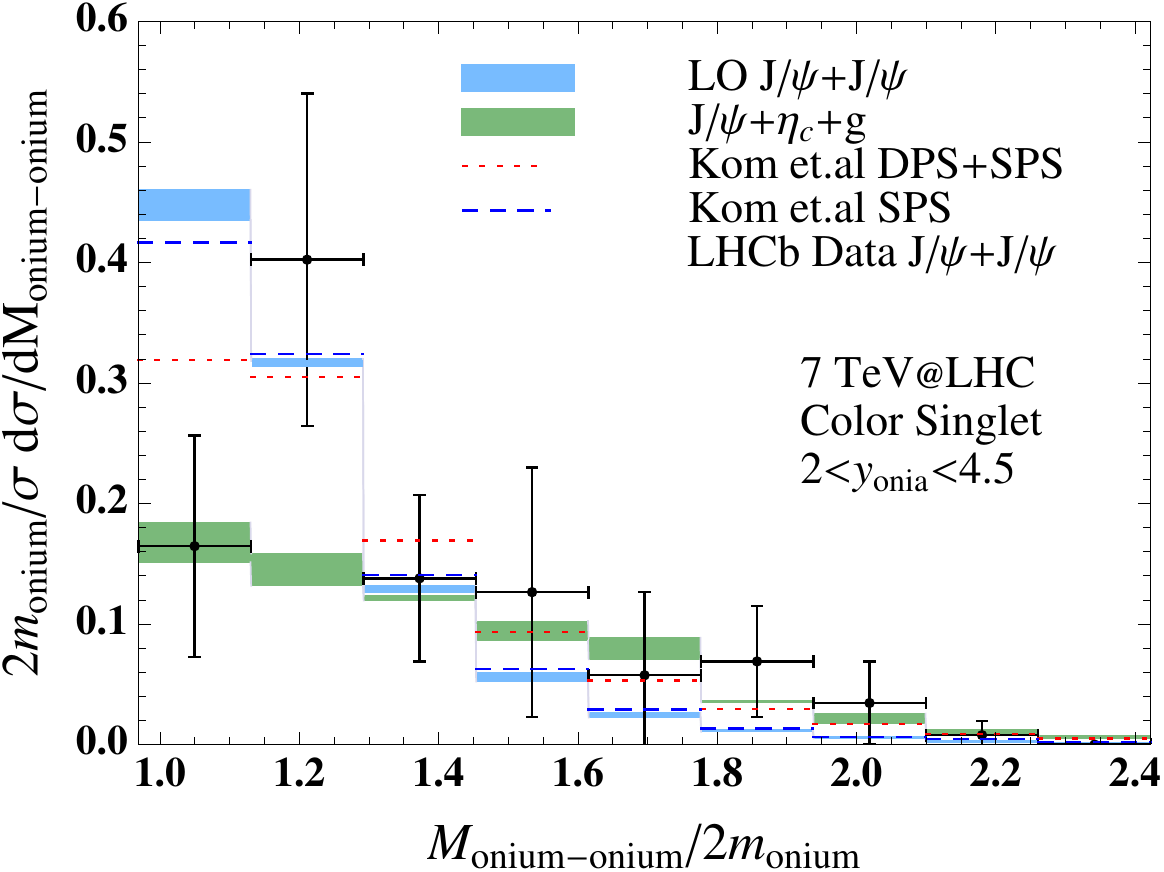}\label{fig:Minv}}\hspace*{1mm}
\subfloat[Polarisation]{\includegraphics[height=4.2cm,clip,draft=false]{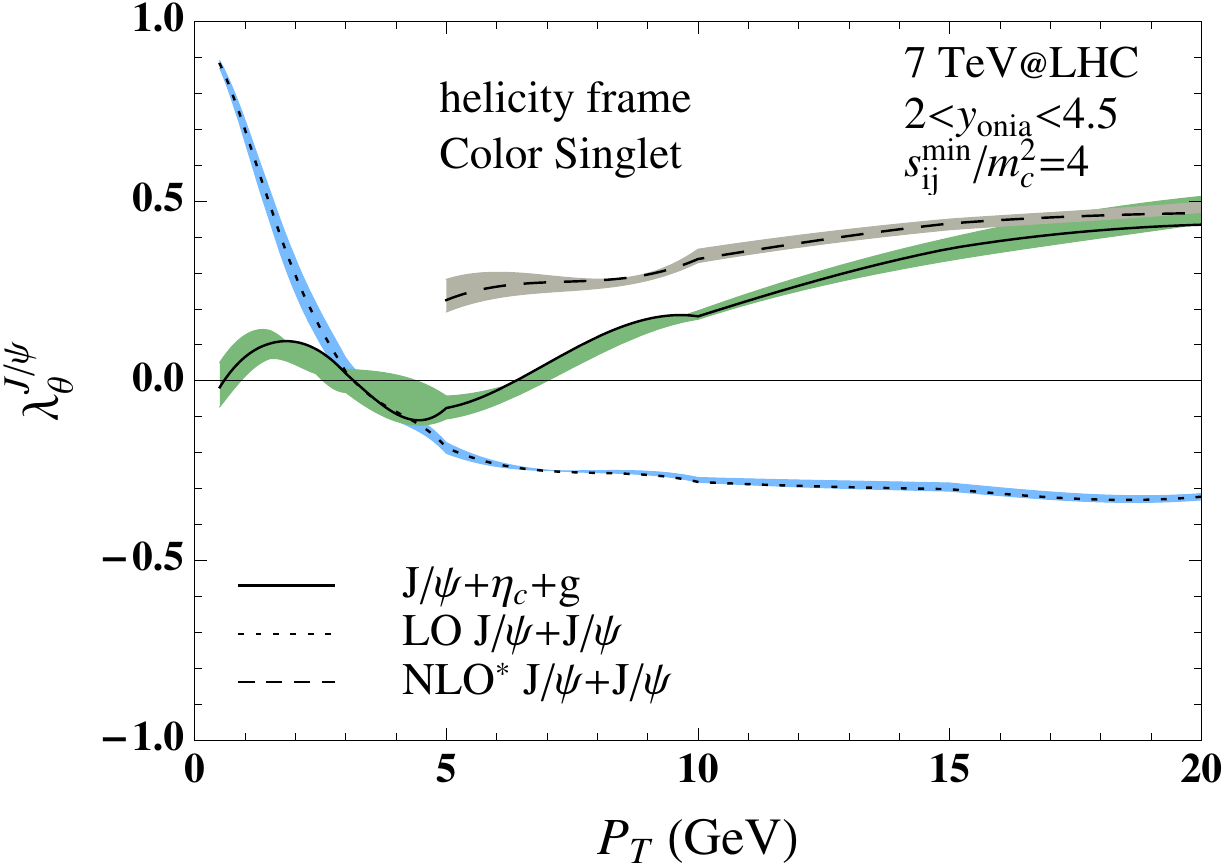}\label{fig:polarisation}}
\caption{$d\sigma/d\Delta \phi$, $d\sigma/d\Delta M_{\rm pair}$  and the polarisation at 7 TeV for $J/\psi+J/\psi$ at LO and NLO$^\star$,  and for $J/\psi+\eta_c+g$.}
\end{center}
\end{figure*}

{\it Cross-section  results .---}
Our LO results for $d\sigma/dP_T$  are shown 
in \cf{fig:dsigdPT} (a)  for $|y|<3$ covered by CMS and ATLAS and (b) for $2.0 < y < 4.5$
covered by LHCb.
At low $P_T$, one observes that the $J/\psi+\eta_c$ yield is about ten times smaller 
than that of double $J/\psi$ for which we note that our results agree with those 
of~\cite{Qiao:2009kg,Berezhnoy:2011xy}. A factor three comes from the spin-state counting 
and another factor from a single power of $\alpha_s$. However, the $P_T$ falloff for 
$J/\psi+\eta_c$ is less strong than for double $J/\psi$ at LO and it is remarkable to note 
that, already at $P_T \simeq 17$ GeV, both yields are similar. One also notes slightly 
larger theoretical uncertainties due to the additional sensitivity on the renormalisation 
scale, $\mu_R$, through 5 instead of 4 powers of $\alpha_s(\mu_R)$.

The next observation is of course the large enhancement between the yield at LO and NLO$^\star$. 
It is absolutely similar to what has been observed for the single $J/\psi$ and $\Upsilon$ 
inclusive production. In these cases, the NLO$^\star$ yield was found to accurately reproduce 
the full NLO yield for $P_T$ as low as a few times $m_Q$. We cannot make such a comparison here 
since the full NLO yield is not known yet. To strengthen our findings, we have separately plotted  
the sensitivity on the IR cut-off $s^{\rm{min}}_{ij}$ on \cf{sijdep}: it is negligible at 
$P_T \gtrsim 10$ GeV and smaller than a factor of two for $P_T \gtrsim 5$ GeV. The 
theoretical uncertainties at NLO$^\star$ shown on Figs. \ref{fig:dsigdPTa} and \ref{fig:dsigdPTb} 
mainly come from the scales and the mass uncertainties. One  notes that it is similar to 
that of $J/\psi+\eta_c+g$, \ie~slightly larger than for  double $J/\psi$ at LO.
 
In order to better understand the $P_T$ scaling of double $J/\psi$ and $J/\psi+\eta_c+g$ 
production, we have also added to \cf{sijdep} the curve for the leading-$P_T$ colour octet (CO) 
contributions, $\so+\so$ (with $\mopb=2\times 10^{-3}$ GeV$^3$), 
whose $d\sigma/dP_T^2$ is expected to scale as $P_T^{-4}$ at high $P_T$ via double 
gluon-fragmentation channels. The $P_T$ scalings are extracted from the ratios of the 
differential cross section in CSM over that of $\so+\so$. As for the inclusive 
single-$J/\psi$ production, $d\sigma/dP_T^2$ for the LO and NLO$^\star$ $J/\psi$-pair 
production scale as $P_T^{-8}$ and $P_T^{-6}$ respectively, while the $J/\psi$-$P_T$ scaling
in $J/\psi+\eta_c+g$  is $P_T^{-7}$. The extra $P_T$ suppression of $J/\psi+\eta_c+g$ 
compared to double $J/\psi$ at $\alpha_s^5$ likely comes from the presence of $t$-channel 
exchange channels for $J/\psi+J/\psi$, absent in $J/\psi+\eta_c+g$. Finally, 
owing to the $P_T^2$ enhancement observed at $\alpha_S^5$, the CSM contributions may 
dominate over the double $J/\psi$ yield with respect to the CO contributions. With the 
parameter value we have chosen, the NLO$^\star$ yield is still ten times the double-CO one 
at $P_T=50$~GeV. Stronger statements would require a careful analysis including the mixed CO+CS
contributions.

{\it Kinematic correlations.---}
Azimuthal correlations and invariant-mass distributions could be useful observables 
to study DPS contributions to associated quarkonium production, whose 
azimuthal distributions should, for instance, be flat. 
Since we are primarily interested in the distribution shapes, we have normalised 
each ones such that their integral between 0 and $\pi$ is one; this has the virtue of 
significantly reducing the uncertainties.

At LO, the $J/\psi$ pair is completely anti-correlated with a peak in 
$\Delta\phi_{J/\psi-J/\psi}$ at $\pi$, \ie~back-to-back. In the presence of additional
final-state particles, the far ``away'' side is not anymore the only populated region. 
This can be observed on \cf{fig:dphi} for $J/\psi+\eta_c+g$, although 
the ``near'' side, $\Delta\phi \simeq 0$, remains barely populated. 
We also checked that the addition of intrinsic $k_T$ for the initial gluons  (see \eg~\cite{Sridhar:1998rt})
creates a natural imbalance (see the lines on~\cf{fig:dphi}) but 
decreases as $\langle k_T\rangle/P_T$. To avoid such $k_T$ effects, it was 
therefore suggested~\cite{Kom:2011bd} to study azimuthal correlations 
with a $P_T$ cut, which we discuss now.

For $J/\psi+\eta_c+g$,  the configurations in which the gluon recoils against 
the pair are not $P_T$-enhanced. On the contrary, for $J/\psi+J/\psi$ at  NLO$^\star$ and
for sufficient $P_T$, the pair recoils against a parton in the $P_T$-enhanced $t$-channel 
exchange topologies. The quarkonia become ``near'' each other and the distribution also peaks
 $\Delta\phi_{J/\psi-J/\psi} \simeq 0$ (see \cf{fig:dphi}). We have also observed that the away-side peak 
decreases for increasing $P_T$. Overall, the introduction of a $P_T$ cut may not suffice
to be able to make clear cut comparison between DPS and SPS distributions.

Let us now study the invariant-mass distributions --normalised to the integrated cross-section-- 
(see~\cf{fig:Minv}) which, in the $J/\psi$-pair 
case, can already be confronted to the LHCb data~\cite{Aaij:2011yc}. At LO for $J/\psi+J/\psi$, we recover
the shape of the SPS results of~\cite{Berezhnoy:2011xy} and~\cite{Kom:2011bd}, which seems 
to agree with the data except for the first bin. We do not entirely share the observation made~\cite{Kom:2011bd}
that the SPS disagrees with the data. We think that uncertainties in the SPS normalisation were
underestimated in~\cite{Kom:2011bd} and it is not clear whether the peak in the second bin --or the dip 
in the first-- is just a fluctuation or a feature of the distribution, more in line with the 
DPS expectations. Forthcoming data will certainly clarify the situation.

Since no $P_T$ cut has been imposed on the data, we cannot 
compare it with the distribution at NLO$^\star$. As a makeshift, we can analyse 
that of $J/\psi+\eta_c$, 
which looks definitively less peaked. This was expected since 
the gluon radiation from the heavy-quark line allows 
for a larger momentum difference between both quarkonia. Such a radiation
definitely appears at NLO, specifically at low $P_T$. Overall, one should be careful
in comparing data with LO predictions only.

{\it Polarisation.---}
We have also found it instructive to evaluate the polarisation, \ie~the 
polar anisotropy $\lambda_\theta$ in the helicity frame (see for instance~\cite{Faccioli:2008dx,Faccioli:2010kd}),
of the $J/\psi$ accompanied by an $\eta_c$ and to compare it
to the polarisation of a $J/\psi$ accompanied by another $J/\psi$. For this comparison to make sense, 
we have therefore evaluated the polarisation in the $J/\psi$ pair case at NLO$^\star$. To our knowledge, this is again 
a new result; it would be confirmed with a full NLO computation. Yet, given the extremely quick disappearance 
of the cut off sensitivity, we believe this result to be reliable, although approximate. 

\cf{fig:polarisation} shows our results in the LHCb kinematics, which we believe to be the only experiment able to measure
polarisation observable for both $J/\psi+J/\psi$ and $J/\psi+\eta_c$ in the future at the LHC\footnote{We note
however that a fixed-target experiment on the LHC beams~\cite{Brodsky:2012vg} would also be an ideal experimental 
set-up to study these processes with high precision~\cite{Lansberg:2012kf}.}. Our results are as follows.
At large $P_T$, the NLO$^\star$ polarisation of a $J/\psi$ produced in pair --transverse-- is quite different than that at LO 
--slightly longitudinal as found in~\cite{Qiao:2009kg}--. This is not a real surprise since, for mid and large $P_T$, the 
NLO$^\star$ is dominated by novel topologies at $\alpha_s^5$. The only known case where such new topologies create $J/\psi$'s with
the same polarisation as the Born topologies is that of $J/\psi+Z$~\cite{Gong:2012ah}. Finally, it is interesting to observe that the NLO$^\star$ results for $J/\psi+J/\psi$ coincide  at high $P_T$
with those for $J/\psi+\eta_c+g$. 

 At low $P_T$, the polarisation for $J/\psi+J/\psi$ at LO and $J/\psi+\eta_c$
seem to converge, then to diverge for $P_T \to 0$.  We are not in position to tell from which $P_T$ 
the polarisation at LO and at NLO$^\star$ are the same since the low $P_T$ region is certainly not the region
where the NLO$^\star$ approximation is the most reliable. 
Finally, we emphasise that these results apply for the production of radially-excited $S$-wave, \ie~$\psi(2S)$ and $\eta_c(2S)$.

{\it Conclusion.---}
We have shown that the LO CSM contributions to direct $J/\psi+\eta_c$  are not small 
despite  its Born contributions being suppressed by one power of $\alpha_s$ compared 
to double $J/\psi$. If the CSM contribution for $J/\psi+\eta_c$ had really been suppressed, 
this process could have been a good probe of colour octet transitions.
Unfortunately, as for many others, it is not so. This is a further illustration 
that conclusions based on the naive power counting on the QCD strong coupling are 
not trustworthy and should always be checked case by case.

In addition, we have evaluated the impact on double $J/\psi$ production 
of the QCD corrections from the real gluon-emission at mid and large $P_T$. We 
have found out that the $P_T$ spectrum is indeed affected in such a way that 
the CSM contributions to double $J/\psi$ might be dominant even at large $P_T$, 
at variance with observations done at LO.

We have also studied  azimuthal correlations and invariant-mass 
distributions. These observables may happen to be useful in determining the 
importance of DPS contributions. Yet, we observed that various effects, such 
as the radiative corrections considered here and 
$k_T$ smearings, make the SPS predictions less clear cut.

Finally, we have analysed polarisation observables. We have showed that QCD corrections 
to double $J/\psi$ production alter the $J/\psi$ spin alignment and that the polarisation 
of direct $J/\psi$ produced with a $\eta_c$ is similar that of a 
$J/\psi$ produced with another $J/\psi$ once these QCD corrections are accounted for.


{\bf Acknowledgements.}
 We thank S. Barsuk, V. Belyaev, S.J.~Brodsky, K.T. Chao, B. Gong, C Lorc\'e, P. Robbe and J.X. Wang for useful discussions. This work is supported in part 
by the France-China Particle Physics Laboratory (FCPPL). H.S.Shao is also supported by the ERC grant 291377, “LHCtheory: Theoretical predictions and analyses of LHC physics: advancing the precision frontier”.



\end{document}